 \documentclass[final,1p,times]{elsarticle}

\usepackage{amssymb}

\journal{Nuclear Physics B}

\begin{document}

\makeatletter
\def\eqnarray{%
 \stepcounter{equation}%
 \let\@currentlabel=\theequation
 \global\@eqnswtrue
 \global\@eqcnt\z@
 \tabskip\@centering
 \let\\=\@eqncr
 $$\halign to \displaywidth\bgroup\@eqnsel\hskip\@centering
 $\displaystyle\tabskip\z@{##}$&\global\@eqcnt\@ne
 \hfil$\displaystyle{{}##{}}$\hfil
 &\global\@eqcnt\tw@$\displaystyle\tabskip\z@{##}$\hfil
 \tabskip\@centering&\llap{##}\tabskip\z@\cr}
\makeatother

\begin{frontmatter}



\title{Constrained WZWN Models on $G/\{S\otimes U(1)^n\}$  \\ and 
Exchange Algebra of  $G$-Primaries }


\author{Shogo Aoyama and Katsuyuki Ishii}

\address{Department of Physics, 
              Shizuoka University 
                Ohya 836, Shizuoka,  
                 Japan}

\begin{abstract}
Consistently constrained WZWN models on $G/\{S\otimes U(1)^n\}$  is given by constraining currents  of the  WZWN models with $G$. Poisson brackets are set up on the light-like plane. Using them we show the Virasoro algebra for the energy-momentum tensor of constrained WZWN models. We find a $G$-primary which satisfies  a classical exchange algebra  in an arbitrary representation of $G$.  The $G$-primary and  the constrained currents are also shown to obey the conformal transformation  with respect to the energy-momentum tensor.  It is checked  that conformal weight  of the  constrained currents is $0$. This  is necessary for the consistency for our formulation of constrained WZWN models.

\end{abstract}




\end{frontmatter}



\section{Introduction}

The AdS/CFT correspondence between the type $IIB$ string theory with the AdS background  and the $N=4$ supersymmetric QCD was studied with great interest.  A remarkable relationship between  the $N=4$ supersymmetric QCD and  a spin-chain system was discovered\cite{Ma,Bei1}. Integrability of the latter system provided a powerful tool for studying the AdS/CFT correspondence. Namely the S-matrix of the latter system  was constructed  by assuming   algebraic structure  with the group symmetry $PSU(2,2|4)$ or its variants\cite{Bei2}. But the construction was  not based on the Poisson structure of some underlying world-sheet theory. 
To explain it, we consider a tensor product chain of quantum operators $\Psi$. We exchange two  of them in a adjacent position, say,  $\Psi(x)$ and $\Psi(y)$.
 Then the S-matrix defines a quantum exchange algebra as
\begin{eqnarray}
S_{xy}\Psi(x)\otimes \Psi(y)=\Psi(y)\otimes \Psi(x).   \label{QEA}
\end{eqnarray} 
The invariant S-matrix under generators $T^A$ of the above symmetry group defined as
\begin{eqnarray}
[T^A\otimes 1+1\otimes T^A, S]=0,   \label{isometry}
\end{eqnarray}
was explicitly given by requiring that 
\begin{eqnarray}
 S_{xy}S_{yx}&=&1,  \label{unitarity}\\
 S_{xy}S_{xz}S_{yz}&=&S_{yz}S_{xz}S_{xy},\quad ({\rm Yang\ Baxter\ equation}).
\label{YBB}
\end{eqnarray}
The R-matrix is defined by $S=P R$ with a permutation operator of the position. Then 
 the relations (\ref{isometry}) and (\ref{YBB}) form the axioms of the quasi-triangular Hopf algebra\cite{Pres}, when supplemented  by two more conditions\cite{Ja,Tori1}: existence of $R^{-1}$  and 
$$
({\cal S}\otimes 1)R=R^{-1},\quad\quad (1\otimes {\cal S}^{-1})R=R^{-1},
$$
with a map ${\cal S}$, called  antipode. 
In physical terms the last relation  is interpreted as  crossing symmetry between  scattering  of particles and   one replaced  by anti-particles\cite{Ja}.  

Let us rewrite the quantum exchange algebra (\ref{QEA}) as
\begin{eqnarray}
R_{yx}\Psi(y)\otimes \Psi(x)=P\Psi(y)\otimes \Psi(x), \label{QEA2}
\end{eqnarray} 
multiplying   the above  permutation operator $P$ on both sides. Suppose the  R-matrix to be a quantum deformation of a certain classical r-matrix
as
$$
R_{xy}= 1+ hr_{xy}+O(h^2),
$$
with an infinitesimal parameter $h$. 
Then (\ref{QEA2}) and (\ref{YBB}) respectively become  a classical exchange algebra
\begin{eqnarray}
 \{\Psi(x)\mathop{,}^\otimes\Psi(y)\}= -hr_{xy}\Psi(x)\otimes \Psi(y), 
\label{CEA}
\end{eqnarray}
and the classical Yang-Baxter equation
\begin{eqnarray}
 [r_{xy},r_{xz}]+[r_{xy},r_{yz}]+[r_{xz},r_{yz}]=0.   \label{YB}
\end{eqnarray}
The quantity in the {\it l.h.s.} of (\ref{CEA}) is a Poisson bracket which was substituted for the commutator $\displaystyle{[\Psi(x)\mathop{,}^\otimes\Psi(y)]}$.  The discovered S-matrix in \cite{Bei2} was of a new type. It did not take the difference form of a spectral parameter\cite{Dri}. Hence the corresponding classical r-matrix was studied with interest as much\cite{Tori2}. But the study was not based on the Poisson structure of some underlying world-sheet theory. 

In the previous paper\cite{Ao} we have studied constrained WZWN models and found
 the classical r-matrix by studying  the Poisson structure.  
 Namely  we have discussed constrained WZWN models with a symmetry group $SL(N)$ based on the coset spaces
\begin{eqnarray}
SL(N)/\{SL(N-M)\otimes SL(M)\otimes U(1)\},\quad\quad SL(N)/\{U(1)^{N-1}\},  \label{SL}
\end{eqnarray}
We have found  $\Psi$ transforming as the fundamental representation  of $SL(N)$, simply called a $SL(N)$ primary . Then it was  shown to satisfy  the classical exchange algebra (\ref{CEA}) and transform as a conformal primary with respect of the energy-momentum tensor. 
The r-matrix of the classical exchange algebra for this case does not take  the classical limit \cite{Tori2} of 
the S-matrix discovered in \cite{Bei2}. It does not take the difference form either\cite{Dri}.  But its existence  naturally leads us to consider constrained WZWN models on the coset spaces (\ref{SL}) as an underlying world-sheet field theory  for  spin-chain systems. Indeed in \cite{Fe} constrained WZWN models  were studied as a generalized Toda theory. The works on the ordinary WZWN models, i.e.,  without  constraints, were done in quest of an underlying world-sheet field theory describing spin-chain systems\cite{Sch}.

 So the aim of this paper is to generalize  the  arguments  to the most extent in this line of thoughts.  In section 2 we formulate constrained WZWN models with any simple group $G$  based on  the coset space 
$$
G/\{S\otimes U(1)^n\}, \quad n=rank\ G-rank\ S,
$$
with an arbitrarily chosen  simple or semi-simple subgroup $S(\subset G)$. 
 It is done by constraining currents of the ordinary WZWN model in a specific way.  
In section 3   Poisson brackets are consistently set up on the light-like plane. The Virasoro algebra  for the energy-momentum tensor is shown by means of them. In section 4 
we  construct a $G$ primary $\Psi$ in arbitrary representations of $G$, which satisfies the classical exchange algebra (\ref{CEA}). In section 5 they are shown to obey the conformal transformation law. We also show  that the constrained currents transform as conformal primaries with weight $0$. It guarantees the consistency of constrained WZWN models.

\section{Constrained WZWN models}

Consider a simple group $G$ on the base of  a subgroup $H$. Generators of $G$ are decomposed into  ones of $H$ and  remaining ones as 
$\{ T_{G/H}, T_H\}$. 
Suppose that $H=S\otimes U(1)^n$ with a simple or semi-simple group $S$ and $n\le\ rank\ G$, and write the generators as 
\begin{eqnarray}
\{T_H\}=\{T_S^i,Q^1,Q^2,\cdots,Q^n \}.       \nonumber
\end{eqnarray}
We define a $Y$-charge such as 
\begin{eqnarray}
T^Y=\sum_{\mu=1}^n Q^\mu.  \nonumber
\end{eqnarray}
Then the coset generators are decomposed as 
\begin{eqnarray}
\{T_{G/H}\}=\{T^{-\alpha}_L,T^\alpha_R\},
      \nonumber
\end{eqnarray}
since they  have positive or negative definite $Y$-charges  
as  
\begin{eqnarray}
 [T^Y, T^{-\alpha}_L ]=-y(\alpha) T^{-\alpha}_R,\quad\quad\quad  
 [T^Y, T^\alpha_R ]=y(\alpha) T^\alpha_R,      \label{Y}
\end{eqnarray}
with $y(\alpha) > 0$. 
They are further decomposed as 
\begin{eqnarray}
\{ T^{-\alpha}_L\}=\{ T^{-\alpha_1}_L,T^{-\alpha_2}_L,\cdots\cdots \},\quad\quad\{ T^{\alpha}_R\}=\{ T^{\alpha_1}_R,T^{\alpha_2}_R,\cdots\cdots \},   \nonumber
\end{eqnarray}
according to the $Y$-charge  $Y_d=y(\alpha_d), d=1,2,\cdots$. This is nothing but an irreducible decomposition of the coset generators under the subgroup $S$. In other words, $\{T^\alpha_R\}$ transform under $S$ as a reducible representation  as
\begin{eqnarray}
\mbox{\boldmath $N$}^{\alpha_1}\oplus \mbox{\boldmath $N$}^{\alpha_2}\oplus\cdots\ ,    \nonumber
\end{eqnarray}
while $\{ T^{-\alpha}_L\}$ as its complex conjugate. Here 
$\alpha_d$ is a set of weights  denoting the  representation  $\mbox{\boldmath $N$}^{\alpha_d}, d=1,2,\cdots$.

A constrained WZWN model with a symmetry group $G$ is  given by the action 
\begin{eqnarray}
S=-{k\over 4\pi}S_{WZWN}-{k\over 2\pi}\int d^2x {\rm Tr}[A_-(g^{-1}\partial_+g-e)] ,    \label{action} 
\end{eqnarray}
in which 
\begin{eqnarray}
g\in G, \quad\quad 
A_-= \sum_{y(\alpha)>0}a^{-\alpha}_-T^\alpha_R\equiv a_-\cdot T_R,
 \nonumber
\end{eqnarray}
and $e$ is a constant matrix in the representation space of $G$.
Here $A_-$ is a gauge field and the action is invariant under local variations 
\begin{eqnarray}
\delta g\longrightarrow gv,\quad\quad 
\delta A_-\longrightarrow -\partial_-v-[A_-,v],  \label{local}
\end{eqnarray}
with an infinitesimal parameter $v= v(x^+,x^-)\cdot T_R$. The action (\ref{action}) is  different from the one of  the ordinary gauged  WZWN models with the gauge field $A_\pm = a_\pm\cdot T_H$, which was given in \cite{Wit}. The latter is invariant under 
  vector-like variations as
\begin{eqnarray}
\delta g\longrightarrow gv-vg,\quad\quad 
\delta A_\pm\longrightarrow -\partial_\pm v-[A_\pm,v].  \label{modlocal}
\end{eqnarray}
with $v=v(x^+,x^-)\cdot T_H$.

The equation of motion for $A_-$ provides the constraint for the current 
\begin{eqnarray}
J^\alpha_+={\rm Tr}[g^{-1}\partial_+gT^\alpha_R]=const..   
 \label{currents}
\end{eqnarray}
We parametrize $g\in G$ by the Gauss decomposition as
 $g= g_Lg_Hg_R$ with 
\begin{eqnarray}
g_L&=&\exp(i{\sum_{y(\alpha)>0}G^{\alpha}T^{-\alpha}_L})\equiv\exp(i{G\cdot T_L}),\nonumber\\
g_R&=&\exp(i\sum_{y(\alpha)>0}F^{-\alpha}T^\alpha_R)\equiv \exp(i{F\cdot T_R}),   \nonumber\\
g_H&=&\exp(i\sum_i\lambda^iT^i_S+i\sum_\mu\lambda^\mu Q^\mu)\equiv
 \exp(i\lambda\cdot T_H). \quad\quad 
  \nonumber
\end{eqnarray}
In this paper we consider $G$ as a compact group so that $T_L^\dag=T_R, T_H^\dag=T_H$. Hence the variables of parametrization  $G^{\alpha},F^{-\alpha},\lambda^i,\lambda^\mu$ are constrained by the unitary condition $g^\dagger g=1$, i.e., 
\begin{eqnarray}
\exp(-i\lambda^* \cdot T_H)\exp({G^*\cdot T_R})\exp({G\cdot T_L})\exp(i\lambda\cdot T_H)=\exp(F^*\cdot T_L)\exp(-F\cdot T_R).  \nonumber
\end{eqnarray}
Here $G^{*-\alpha}, F^{*\alpha}, \lambda^{*i},\lambda^{*\mu}$ are complex conjugates of $G^{\alpha}, F^{-\alpha}, \lambda^i,\lambda^\mu$. 
We choose the  gauge $F^{-\alpha}=F^{*\alpha}=0$ and denote the remaining 
variables by $\cal G$. 
The gauge-fixed transformation of $g({\cal G})=g_L(G^\alpha)g_H(\lambda^i,\lambda^\mu)$ is given by 
\begin{eqnarray}
g({\cal G})\longrightarrow e^{i\epsilon\cdot T}g({\cal G})e^{-u}=g({\cal G}'),  \label{transf}
\end{eqnarray}
with a unitary element $e^{i\epsilon\cdot T}\in G$  and an appropriate compensator $e^{-u}$ of which generator is 
\begin{eqnarray}
u=\sum_{y(\alpha)>0} u^{-\alpha}T^\alpha_R.   \label{u}
\end{eqnarray}
This is nothing but  a symmetry transformation on the coset space $G/\hat H$ with the complex subgroup $\hat H$ generated by $T^{-\alpha}_L$ and $T_H$. Soon later we show that the  action (\ref{action}) is indeed invariant by this transformation even after the gauge-fixing. This symmetry plays a crucial role in the subsequent discussions. However  
$g({\cal G}')$ in (\ref{transf}) loses the unitarity because the  compensator never satisfies $e^ue^{u^\dagger}=1$. Hence the gauge-fixed action is not real. It is a natural consequence because the action (\ref{action}) is complex from the beginning.  At this point our WZWN models  
differ from  $SL(N)$ WZWN models in \cite{Ao}, where $g=g_Lg_H$ merely satisfies $\det g=1$ and the gauge-fixing does not break the reality of the action. Losing  the unitarity by the gauge-fixing  we have no reason to impose a further constraint as $(J^{\alpha}_+)^*=J^{-\alpha}_+$. Indeed such a constraint holds no longer since  we have $J^{-\alpha}_+\equiv {\rm Tr}[g^{-1}({\cal G})\partial_+g({\cal G})T^{-\alpha}_L]=0$. 

 Let us assume that $e^{i\epsilon\cdot T}$ depends on $x^-$ alone. Then 
the transformation (\ref{transf}) induces 
\begin{eqnarray}
\delta(g^{-1}\partial_+g)=-\partial_+u-[g^{-1}\partial_+g,u]. 
\label{gauge-transf}
\end{eqnarray}
We understand the  transformation (\ref{transf}) with this assumption hereinafter all the time. 
Under this the currents defined in
 (\ref{currents}) transform as 
\begin{eqnarray}
\delta J^\alpha_+=-{\rm Tr}(g^{-1}\partial_+g[u,T^\alpha_R]),
  \label{delta J}
\end{eqnarray}
by the constraints (\ref{currents}). We put the $Y$-charges of the  generators $T^\alpha_R$ in an increasing order  as  $0<Y_1<Y_2<\cdots$. We take  the  constraints (\ref{currents})
 on $J^\alpha_+$ in the specific form 
\begin{eqnarray}
 {\rm Tr}[g^{-1}\partial_+gT^\alpha_R]=\left\{
\begin{array}{rl}
const, &\quad \mbox{for  $Y_1=y(\alpha)$ },  \\
    &     \\
  0,  & \quad \mbox{for other $y(\alpha)$'s}. 
\end{array}\right.   \label{constraint}
\end{eqnarray}
Then the variation (\ref{delta J}) is vanishing because $u$ takes the form (\ref{u}) and $[u,T^\alpha_R]$ is valued in the Lie algebra of $G$ with the  $Y$-charge larger than the lowest $Y_1$. This guarantees the self-consistency of the constraints (\ref{constraint}). 
   Next we consider the energy-momentum tensor $T_{++}$. The one of the naive form is no longer invariant under the gauge-fixed transformation (\ref{transf}), so that we improve it as 
\begin{eqnarray}
 T_{++}=k\Big({1\over 2}{\rm Tr}[(g^{-1}\partial_+g)^2]+{1\over Y_1}\partial_+{\rm Tr}[g^{-1}\partial_+ gT^Y]\Big).    \label{T}
\end{eqnarray}
This is invariant under the gauge-fixed transformation  as
\begin{eqnarray}
\delta T_{++}&=& -k{\rm Tr}[(g^{-1}\partial_+g)\partial_+u] -{1\over Y_1}k\partial_+{\rm Tr}((g^{-1}\partial_+g)[u,T^Y]) \nonumber \\
  &=& -k{\rm Tr}[(g^{-1}\partial_+g)\sum_{\alpha_1} \partial_+u^{-\alpha_1}T^{\alpha_1}_R]
  -k\partial_+{1\over Y_1}{\rm Tr}((g^{-1}\partial_+g)[\sum_{\alpha_1}u^{-\alpha_1}T^{\alpha_1}_R,T^Y])=0,   \nonumber
\end{eqnarray}
using (\ref{gauge-transf}), (\ref{u}) and the above constraints. Similarly we can show that the 
 action  (\ref{action})  is invariant under the transformation (\ref{transf}), even after the gauge is fixed.

We study the gauge-fixed transformation (\ref{transf}) at  two steps such as 
\begin{eqnarray}
g_L(G^\alpha)&\longrightarrow& e^{i\epsilon\cdot T}g_L(G^\alpha) e^{-i\rho\cdot \hat H}=g_L(G'^\alpha),   \label{1st} \\
g_H(\lambda^i,\lambda^\mu)&\longrightarrow& e^{i\rho\cdot \hat H}g_H(\lambda^i,\lambda^\mu)e^{-u}=g_0(\lambda'^i,\lambda'^\mu),  \label{2nd}
\end{eqnarray}
in which 
\begin{eqnarray}
\epsilon\cdot T &=&\sum_{y(\alpha)>0}\epsilon_R^\alpha T_L^{-\alpha}+
\sum_{y(\alpha)>0}\epsilon_L^{-\alpha}T_R^\alpha+
\sum_i \epsilon_H^iT_H^i +\sum_\mu \epsilon^\mu Q^\mu,
   \nonumber \\
\rho\cdot \hat H &=& \sum_{y(\alpha)>0}\rho_L^{-\alpha}T_R^\alpha+
\sum_i \rho_H^iT_H^i +\sum_\mu \rho^\mu Q^\mu.
  \nonumber
\end{eqnarray}
$\rho\cdot \hat H$ is the generator of the compensator for the transformation at the first step. Both transformations are written in the infinitesimal forms 
\begin{eqnarray}
\delta G^\alpha&=&{\rm Tr}[(\epsilon\cdot T g_L-g_L\rho\cdot \hat H)T_R^\alpha] \equiv t_{AB}\epsilon^A\delta^B G^\alpha, \nonumber\\
\delta \lambda^i&=&{\rm Tr}[(\rho\cdot \hat H g_H-g_H u)T_S^i]
\equiv t_{AB}\epsilon^A\delta^B \lambda^i, \nonumber\\ 
\delta \lambda^\mu&=&{\rm Tr}[(\rho\cdot \hat H \lambda^\mu-\lambda^\mu u)Q^\mu]
\equiv t_{AB}\epsilon^A\delta^B \lambda^\mu,    \nonumber
\end{eqnarray}
which  define the Killing vectors $\delta^A {\cal G}, A=1,2,\cdots,dim\ G$, which realize the symmetry of the coset space $G/\{S\otimes U(1)^n\}$. The variables $G^\alpha$ are coordinates of the coset space, while $\lambda^i,\lambda^\mu$ auxiliary. An important point is that owing to $\delta J^\alpha_+=0$  the constraints (\ref{constraint}) do not reduce  the symmetry of the coset space.

\section{The  Virasoro transformation}

We shall set up Poisson brackets for the group variables ${\cal G}^I=(G^\alpha,\lambda^i,\lambda^\mu)$. The guiding principle to do this is that they satisfy the Jacobi identities and are able to reproduce the Virasoro algebra for the energy-momentum tensor (\ref{T}). They are given by 
\begin{eqnarray}
&\ & \{{\cal G}^I(x)\mathop{,}^\otimes {\cal G}^J(y)\}   \nonumber \\
&\ &\quad\quad ={2\pi\over k }\Big[\theta(x-y)t_{AB}^+ \delta^A{\cal G}^I(x)\otimes \delta^B{\cal G}^J(y) -\theta(y-x)t_{AB}^+ \delta^A{\cal G}^J(y)\otimes \delta^B{\cal G}^I(x)\Big], \quad\     \label{Poisson}
\end{eqnarray}
at $x^-=y^-$. 
The Killing vectors $\delta^A {\cal G}$ satisfy the Lie algebra of $G$
\begin{eqnarray}
\delta^{[A}\delta^{B]}{\cal G}(x)=f^{AB}_{\ \ \ C}\delta^C{\cal G}(x),  \label{Lie1}
\end{eqnarray}
by the construction. $t_{AB}^+$ is a modified Killing metric such as defining the classical r-matrix 
\begin{eqnarray}
r^{\pm}&=&t_{AB}^{\pm}T^A\otimes T^B    \nonumber \\
&=&
\sum_{y(\alpha)>0}T_R^{\alpha}\otimes T_L^{-\alpha}+
\sum_{\hat\alpha>0} T_{SR}^{\hat\alpha}\otimes T_{SL}^{-\hat\alpha} 
-
\sum_{y(\alpha)>0}T_L^{-\alpha}\otimes T_R^{\alpha}-
\sum_{\hat\alpha>0} T_{SL}^{-\hat\alpha}\otimes T_{SR}^{\hat\alpha} \nonumber\\
&\ & \hspace{4cm} \pm t_{AB}T^A\otimes T^B,
   \label{r-matrix}
\end{eqnarray}
with $t_{AB}^+=-t_{BA}^-$. For $r^+$ it reads 
\begin{eqnarray}
r^+=2\sum_{y(\alpha)>0}T_R^{\alpha}\otimes T_L^{-\alpha}+
2\sum_{\hat\alpha>0} T_{SR}^{\hat\alpha}\otimes T_{SL}^{-\hat\alpha} 
+
\sum_{\hat \mu=1}^{dim S} Q^{\hat\mu}\otimes Q^{\hat\mu}
+\sum_{\mu=1}^n Q^\mu\otimes Q^\mu.   \nonumber
\end{eqnarray}
It satisfies the classical Yang-Baxter equation (\ref{YB}). Here  the generators  of the subgroup $S$ were further decomposed in the Cartan-Weyl basis as 
$$
\{T_S^i\}= \{T_{SL}^{\ -\hat \alpha},T_{SR}^{\ \ \hat \alpha},Q^{\hat\mu}\}, \quad \hat\mu=1,2,\cdots,rank\ S,
$$
with the normalization  ${\rm Tr}T^AT^B={1\over 2}\delta^{AB}\equiv{1\over 2}t^{AB}$.\footnote{ Therefore we have $t^{AB}=t_{AB}$. This definition of the Killing metric is in  accord  with the one in \cite{Ao}. }
The Jacobi identities for the Poisson brackets (\ref{Poisson})  can be 
shown in the same way as in the previous paper\cite{Ao}. Namely the proof there  can be straightforwardly generalized to the case in this paper. 
The Virasoro algebra for the energy-momentum tensor (\ref{T}) follows by means of these Poisson brackets. It can be also shown in the same way as in \cite{Ao}. 
 Here it suffices  to outline  the demonstration of the  Virasoro algebra referring the details to \cite{Ao}.  At $x^-=y^-$ we have 
\begin{eqnarray}
&\ &\{T_{++}(x) \mathop{,}^\otimes T_{++}(y)\}     \nonumber\\
&\ &\quad =k\Bigg({\rm Tr}[(\partial_x gg^{-1})\partial_x(\{g\mathop{,}^\otimes T_{++}(y)\}g^{-1})]   
+ {1\over Y_1}\partial_x{\rm Tr}[\partial_x(\{g\mathop{,}^\otimes T_{++}(y)\}g^{-1})gT^Yg^{-1}]\Bigg), \quad\quad  \label{TT}
\end{eqnarray}
with the help of the formula for a generic variation
\begin{eqnarray}
\delta(g^{-1}\partial_xg)=g^{-1}\partial_x(\delta g g^{-1})g.  \label{variation}\end{eqnarray}
We further calculate the Poisson bracket  $\displaystyle{\{g\mathop{,}^\otimes T_{++}(y)\}}$ in the {\it r.h.s.} to find 
\begin{eqnarray}
&\ & \{g(x)\mathop{,}^\otimes T_{++}(y)\}  \nonumber\\
&\ &\quad\quad 
= k\Bigg({\rm Tr}[\partial_y(\{g(x)\mathop{,}^\otimes g\}g^{-1})
(\partial_y gg^{-1})]+{1\over Y_1}\partial_y{\rm Tr}[\partial_y(\{g(x)\mathop{,}^\otimes g\}g^{-1})gT^Yg^{-1}]\Bigg). \quad\quad \label{gg}
\end{eqnarray}
Finally we have to calculate the Poisson bracket $\displaystyle{\{g(x)\mathop{,}^\otimes g\}}$. To this end we have recourse to the formula
\begin{eqnarray}
\{g(x)\mathop{,}^\otimes g(y)\} =
{\partial g(x)\over \partial {\cal G}^I(x)}\{{\cal G}^I(x)\mathop{,}^\otimes {\cal G}^J(y)\}{\partial g(y)\over \partial {\cal G}^J(y)}.   \label{defPoi}
\end{eqnarray}
By means of the Poisson brackets (\ref{Poisson}) it reads
\begin{eqnarray}
\{g(x)\mathop{,}^\otimes g(y)\} = {2\pi\over k}\Big[\theta(x-y)t_{AB}^+\delta^Ag(x)\otimes \delta^Bg(y)
-\theta(y-x)t_{AB}^+\delta^Ag(y)\otimes\delta^Bg(x)\Big].  \nonumber\\   
\label{Poi} 
\end{eqnarray}
Plug this Poisson bracket into the {\it r.h.s.} of (\ref{gg}). First of all note that  (\ref{gg}) may be put into a simplified form 
\begin{eqnarray}
&\ & \{g(x)\mathop{,}^\otimes T_{++}(y)\}  \nonumber\\
&\ &\quad\quad 
= k\Bigg({\rm Tr}[\partial_y(\{g(x)\mathop{,}^\otimes g\}g^{-1})
(\partial_y gg^{-1})]+{1\over Y_1}\partial_y^2{\rm Tr}[\{g(x)\mathop{,}^\otimes g\}g^{-1}T^Y]\Bigg), \quad\quad \label{gg'}
\end{eqnarray}
as follows. The quantity $\delta gg^{-1}$ by the transformation (\ref{transf})   is valued in the subalgebra formed by $\{T^{-\alpha}_L, T_H\}$.  
So is the Poisson bracket $\displaystyle{\{g(x)\mathop{,}^\otimes g\}g^{-1}}$ from (\ref{Poi}). 
Therefore we can  simplify  the second term of (\ref{gg}) 
as 
\begin{eqnarray}
\partial_y{\rm Tr}[\partial_y(\{g(x)\mathop{,}^\otimes g\}g^{-1})gT^Yg^{-1}]
&=& \partial_y{\rm Tr}[\partial_y(\{g(x)\mathop{,}^\otimes g\}g^{-1})g_LT^Yg_L^{-1}]   \nonumber\\
&=&\partial_y^2{\rm Tr}[\{g(x)\mathop{,}^\otimes g\}g^{-1}T^Y],  \label{simplification}
\end{eqnarray}
to find (\ref{gg'}).  Owing to $\delta^B T_{++}(y)=0$ the {\it r.h.s.} of (\ref{gg'}) is vanishing except when the derivative $\partial_y$ acts on the step functions $\theta(x-y)$ and $\theta(y-x)$ in calculating $\displaystyle{\partial_y\{g(x)\mathop{,}^\otimes g\}}$. Hence picking up both contributions we get 
\begin{eqnarray}
\{g(x) \mathop{,}^\otimes T_{++}(y)\} &=& 4\pi\Bigg(\partial_y\theta(x-y)t_{AB}\delta^Ag(x)\otimes {\rm Tr}[(\delta^Bgg^{-1}) (\partial_ygg^{-1})] \nonumber\\
&\ & \hspace{1cm}+{1\over Y_1} \partial_y^2\theta(x-y)t_{AB}\delta^Ag(x)\otimes {\rm Tr}[(\delta^Bgg^{-1})T^Y]   \nonumber\\
&\ & \hspace{2cm}+{2\over Y_1} \partial_y\theta(x-y)t_{AB}\delta^Ag(x)\otimes \partial_y{\rm Tr}[(\delta^Bgg^{-1})T^Y]\ \Bigg).\ \ \ \   \label{gT'}
\end{eqnarray}
Here note that  $t_{AB}^+$ could be  changed  to the usual Killing metric $t_{AB}$. 
Finally we evaluate  the Poisson bracket (\ref{TT}) by plugging  this expression for $\displaystyle{\{g\mathop{,}^\otimes T_{++}(y)\}} $. For the rest of the calculations the reader may refer to \cite{Ao}. 
It leads us to  
 the Virasoro transformation
\begin{eqnarray}
&\ &{1\over 2\pi}\int dx\eta(x) \{T_{++}(x) \mathop{,}^\otimes T_{++}(y)\}
  \nonumber\\
&\ & \hspace{2cm} =
\eta(y)\partial_yT_{++}(y)+2\Big(\partial_y\eta(y)\Big)T_{++}(y) -{k\over Y_1} {\rm Tr}[T^YT^Y]\partial_y^3\eta(y),
 \label{Virasoroo}
\end{eqnarray}
with an infinitesimal parameter $\eta(y)$. Here use was made of the formula 
\begin{eqnarray}
{\rm Tr}O_1O_2=2t_{AB}({\rm Tr}O_1T^A)({\rm Tr}O_2T^B),    \label{trace}
\end{eqnarray}
for  quantities $O_1$ and $O_2$ valued in the Lie algebra of $G$, with the normalization  given below (\ref{r-matrix}).

So far our arguments are irrelevant to the representation of $G$. Let us choose it to be  an $N$-dimensional irreducible representation. Under the subgroup $S$ it is decomposed into irreducible ones  as 
\begin{eqnarray}
\mbox{\boldmath $N$}= \mbox{\boldmath $N$}^{w_1}\oplus \mbox{\boldmath $N$}^{w_2}\oplus\cdots\oplus\mbox{\boldmath $N$}^{w_{a-1}}\oplus\mbox{\boldmath $N$}^{w_a}.   
\label{decomp}
\end{eqnarray}
Here $w_d,d=1,2,\cdots,a$, is a set of weights denoting the $N_d$-plet representation.
An element $g\in G$ is represented  by a $N\times N$ matrix $D(g)$. For $g_H \in H$ it is decomposed  as
\newfont{\bg}{cmr10 scaled\magstep4}
\newcommand{\bigzerol}{\smash{\hbox{\bg 0}}}
\newcommand{\bigzerou}{%
 \smash{\lower1.7ex\hbox{\bg 0}}}

\begin{eqnarray}
D(g_H)=\left(
\begin{array}{cccc}
\noalign{\vskip0.2cm}D_{w_1}(g_H)  &\hspace{-0.2cm}0&  \cdots    &\hspace{-0.2cm}0 \\
\noalign{\vskip0.2cm} 
0 &\hspace{-0.2cm} D_{w_2}(g_H) & \cdots   &\hspace{-0.2cm}0      \\
\noalign{\vskip0.2cm}
 \vdots  &\hspace{-0.2cm}\vdots     &  \ddots  &\hspace{-0.2cm} \vdots     \\
\noalign{\vskip0.2cm}
  0   &\hspace{-0.2cm}0    &  \cdots    &  D_{w_a}(g_H)     \\
\noalign{\vskip0.2cm}
\end{array}
\right),
\end{eqnarray}
in which  $D_{w_d}(g_H), d=1,2,\cdots, a$, is a $N_d\times N_d$ matrix. Each $\mbox{\boldmath $N$}^{w_d}$ have a definite $Y$-charge $y_d$. Putting them in order as
 $y_1<y_2<\cdots <y_{a-1}<y_a$ we have 
\begin{eqnarray}
D(T^Y) =\sum_{\mu=1}^n D(Q^\mu)=\left(
\begin{array}{cccc}
\noalign{\vskip0.2cm}
 (y_1)_{\scriptscriptstyle{N_1\times N_1}}  &\hspace{-0.6cm}0&\hspace{-0.3cm}\cdots  &\hspace{-0.2cm}0  \\
\noalign{\vskip0.2cm} 
0 &\hspace{-0.6cm} (y_2)_{\scriptscriptstyle{N_2\times N_2}}    &\hspace{-0.3cm} \cdots&\hspace{-0.2cm}0     \\
\noalign{\vskip0.2cm}
 \vdots  &\hspace{-0.6cm}  \vdots   &\hspace{-0.3cm}  \ddots  &\hspace{-0.2cm} \vdots     \\
\noalign{\vskip0.2cm}
  0   &\hspace{-0.6cm}0     &\hspace{-0.3cm}  \cdots    &\hspace{-0.2cm} (y_{a})_{\scriptscriptstyle{N_{a}\times N_{a}}}    \\
\noalign{\vskip0.2cm}
\end{array}
\right), \hspace{2.2cm}   \label{y}
\end{eqnarray}
with  $(y_d)_{N_a\times N_a}\equiv y_d\cdot (1)_{N_d\times N_d}$. 
The central charge of (\ref{Virasoroo}) is given by using this as
\begin{eqnarray}
c=12{k\over Y_1}{\rm Tr}[T^YT^Y]=12{k\over Y_1}{\rm Tr}[D(T^Y)D(T^Y)]=12{k\over Y_1}\sum_{d=1}^a N_d y_d^2.      \label{central}
\end{eqnarray}

\section{The classical exchange algebra}

We are in a position to discuss the classical exchange algebra in arbitrary representation of $G$, say, $\mbox{\boldmath $N$}$. We shall find a primary $\Psi(x)$  which is a composite of ${\cal G}(=\{G(x),\lambda^i(x),\lambda^\mu(x)\})$ and satisfies 
\begin{eqnarray}
\{\Psi(x)\mathop{,}^\otimes \Psi(y)\}={2\pi\over k}[\theta(x-y)\rho(r^+) + \theta(y-x)\rho(r^-)]\Psi(x)\otimes\Psi(y).   \label{CEA'}
\end{eqnarray}
Here $D(r^\pm)$ are  r-matrices given by (\ref{r-matrix}) in the representation $\mbox{\boldmath $N$}$. By the identification 
\begin{eqnarray}
r_{xy}\equiv {2\pi\over k}[\theta(x-y)\rho(r^+) + \theta(y-x)\rho(r^-)],
 \label{rxy}
\end{eqnarray}
this takes the form of the classical exchange algebra (\ref{CEA}). 
From $t_{AB}^+=-t_{BA}^-$ note that $Pr^-_{xy}=-r^+_{yx}$ with the permutation operator defined by (\ref{QEA2}). 
For the above purpose it suffices to find a $N$-dimensional vector which linearly transforms by the gauge-fixed transformation (\ref{transf}) as 
\begin{eqnarray}
\delta \Psi(x)=D(\epsilon\cdot T)\Psi(x), \label{linitransf}
\end{eqnarray}
which is called $G$-primary. 
When we write  $g=g_Lg_H$ in the  representation $\mbox{\boldmath $N$}$ as
\begin{eqnarray}
D(g)=\left(
\begin{array}{cccc}
\noalign{\vskip0.2cm}
(g)_{\scriptscriptstyle{N_1\times N_1}}  & (g)_{\scriptscriptstyle{N_1\times N_2}}  & \cdots    & (g)_{\scriptscriptstyle{N_1\times N_a}} \\
\noalign{\vskip0.2cm} 
(g)_{\scriptscriptstyle{N_2\times N_1}} &(g)_{\scriptscriptstyle{N_2\times N_2}} & \cdots& (g)_{\scriptscriptstyle{N_2\times N_a}}    \\
\noalign{\vskip0.2cm}
 \vdots  &\hspace{-0.2cm}  \vdots   &  \ddots  &\hspace{-0.2cm} \vdots    \\
\noalign{\vskip0.2cm}
 (g)_{\scriptscriptstyle{N_a\times N_1}}     &(g)_{\scriptscriptstyle{N_a\times N_2}}     &  \cdots    &  (g)_{\scriptscriptstyle{N_{a}\times N_{a}} }    \\
\noalign{\vskip0.2cm}
\end{array}
\right).    \label{rho}
\end{eqnarray}
Then such a $G$-primary is given by 
\begin{eqnarray}
\Psi(x)=\left(
\begin{array}{c}
\noalign{\vskip0.2cm}
(g)_{\scriptscriptstyle{N_1\times N_a}}   \\
\noalign{\vskip0.2cm} 
(g)_{\scriptscriptstyle{N_2\times N_a}}   \\
\noalign{\vskip0.2cm}
 \vdots    \\
\noalign{\vskip0.2cm}
 (g)_{\scriptscriptstyle{N_a\times N_a}}     \\
\noalign{\vskip0.2cm}
\end{array}
\right).    \label{Psi}
\end{eqnarray}
The linear transformation law (\ref{linitransf}) is shown as follows. 
The gauge-fixed transformation (\ref{transf}) is written in the infinitesimal form 
\begin{eqnarray}
D(\delta g)=D(\epsilon\cdot Tg)-D(gu).    \label{infini}
\end{eqnarray}
$D(g)$ transforms as a tensor product $\mbox{\boldmath $N$}\otimes \overline{\mbox{\boldmath $N$}}$, which is decomposed by (\ref{decomp}) as 
$$
\mbox{\boldmath $N$}\otimes \overline{\mbox{\boldmath $N$}}
= [\mbox{\boldmath $N$}^{w_1}\oplus \mbox{\boldmath $N$}^{w_2}\oplus\cdots\oplus\mbox{\boldmath $N$}^{w_{a-1}}\oplus\mbox{\boldmath $N$}^{w_a}]\otimes
    [\overline{\mbox{\boldmath $N$}}^{-w_1}\oplus \overline{\mbox{\boldmath $N$}}^{-w_2}\oplus\cdots\oplus\overline{\mbox{\boldmath $N$}}^{-w_{a-1}}\oplus\overline{\mbox{\boldmath $N$}}^{-w_a}].
$$
So does $D(g)D(u)$. However $u$ is given by (\ref{u}) and $D(u)$ transforms as
$$
\mbox{\boldmath $N$}^{\alpha_1}\oplus \mbox{\boldmath $N$}^{\alpha_2}\oplus\cdots\oplus\mbox{\boldmath $N$}^{\alpha_{n-1}}\oplus\mbox{\boldmath $N$}^{\alpha_n}.
$$
The matrix multiplication $D(g)D(u)$ implies that the tensor product 
\begin{eqnarray}
\mbox{\boldmath $N$}\otimes \Big(\overline{\mbox{\boldmath $N$}}
\otimes 
 [\mbox{\boldmath $N$}^{\alpha_1}\oplus \mbox{\boldmath $N$}^{\alpha_2}\oplus\cdots\oplus\mbox{\boldmath $N$}^{\alpha_{n-1}}\oplus\mbox{\boldmath $N$}^{\alpha_n}]\Big),     \nonumber
\end{eqnarray}
is decomposed into a tensor product
\begin{eqnarray}
\mbox{\boldmath $N$}\otimes[\overline{\mbox{\boldmath $N$}}^{-w_1}\oplus \overline{\mbox{\boldmath $N$}}^{-w_2}\oplus\cdots\oplus\overline{\mbox{\boldmath $N$}}^{-w_{a-1}}\oplus\overline{\mbox{\boldmath $N$}}^{-w_a}].
 \nonumber 
\end{eqnarray}
Here  the component $\mbox{\boldmath $N$}\otimes \overline{\mbox{\boldmath $N$}}^{-w_a}$ with the lowest $Y$-charge can never appear because of the positivity of the $Y$-charge of $D(u)$.  Therefore multiplying $D(g)$ by $D(u)$  in 
the {\it r.h.s.} of (\ref{infini}) annihilates  the far right column vectors of $D(g)$, denoted as (\ref{Psi}).   Hence we have the transformation law (\ref{linitransf}).

\section{Conformal transformation of $\Psi$ and $J^\alpha_+$}

In the previous section we have shown that the $G$-primary $\Psi$ linearly transforms as (\ref{linitransf}) under the gauge-fixed transformation (\ref{transf})  and satisfies the classical exchange algebra (\ref{CEA'}). In this section we will examine its transformation property with respect to the energy-momentum tensor $T_{++}$. We show that it indeed transforms as a conformal primary such that 
\begin{eqnarray}
{1\over 2\pi}\int dx\ \eta(x)\{T_{++}(x)\mathop{,}^\otimes \Psi(y) \}=\eta(y)\partial_y\Psi(y) + \Bigg((y_a)_{N_a\times N_a}\sum_{d=1}^a N_d y_d^2\Bigg)\Big(\partial_y\eta(y)\Big) \Psi.    \label{Psiconf}
\end{eqnarray}
In section 2 
we have also seen that the constrained currents $J^\alpha_+$, given by (\ref{currents}), are invariant by the gauge-fixed transformation (\ref{transf}). 
For imposing the constraints (\ref{currents}) consistently it is crucially important that the constrained currents are  conformal primaries with weight $0$ as
\begin{eqnarray}
{1\over 2\pi}\int dx\ \eta(x)\{T_{++}(x)\mathop{,}^\otimes J_{+}^{\alpha}(y) \}=\eta(y)\partial_yJ_{+}^{\alpha}(y).    \label{J+}
\end{eqnarray}
We shall show these conformal transformation laws in this section.
 To this end we give here the useful formulae in connection with the compensator $e^{-u}$ for the gauge-fixed transformation (\ref{transf})
\begin{eqnarray}
{\rm Tr}
[gug^{-1}(\partial_ygg^{-1})] &=& {1\over Y_1}\partial_y{\rm Tr}[(\delta gg^{-1})T^Y],  \label{formula} \\
{\rm Tr}[T_L^{-\alpha}g^{-1}T_R^\beta g]&=& u^{\beta,-\alpha},   \label{u''}
\end{eqnarray}
with $g=g_Lg_H$. 
The first formula can be shown as follows. Note that 
$$
\delta{\rm Tr}[(g^{-1}\partial_yg)T^Y]=-{\rm Tr}[(g^{-1}\partial_yg)[u,T^Y]]= Y_1{\rm Tr}[(g^{-1}\partial_yg)u],
$$
by the successive use of (\ref{gauge-transf}), (\ref{u}), (\ref{Y}) and the constraints (\ref{constraint}) for $J_{+}^\alpha$. This variation can be calculated also  as
$$
\delta{\rm Tr}[(g^{-1}\partial_yg)T^Y]=
{\rm Tr}[\partial_y(\delta gg^{-1})gT^Yg^{-1}]=
\partial_y{\rm Tr}[(\delta gg^{-1})T^Y], 
$$
by using (\ref{variation}) and the same argument as for (\ref{simplification}). 
Equating both variations yields the formula (\ref{formula}). 
The second formula can be also shown by the argument for (\ref{simplification}). Namely we have 
$$
0={\rm Tr}[T_L^{-\alpha}g^{-1}(\delta gg^{-1})g]=
{\rm Tr}[T_L^{-\alpha}g^{-1}(\epsilon\cdot T)g]-\sum_{y(\beta)>0}\epsilon_L^{-\beta}u^{\beta,-\alpha},
$$
in which (\ref{infini}) was substituted for $\delta gg^{-1}$ and (\ref{u}) was 
further decomposed as
\begin{eqnarray}
u=\sum_{y(\gamma)>0} u^{-\gamma}T^\gamma_R=\sum_{y(\beta)>0,\ y(\gamma)>0}\epsilon_L^{-\beta}u^{\beta,-\gamma}T_R^\gamma.  \label{u'}  
\end{eqnarray}
This gives the relation (\ref{u''}). 

\vspace{0.5cm}

\noindent
{\bf i. Conformal transformation of $\Psi$}

Using the Poisson bracket (\ref{gT'}) 
we have 
\begin{eqnarray}
\{\Psi(x) \mathop{,}^\otimes T_{++}(y)\} &=& 4\pi\Bigg(\partial_y\theta(x-y)t_{AB}\delta^A\Psi(x)\otimes {\rm Tr}[(\delta^Bgg^{-1}) (\partial_ygg^{-1})] \nonumber\\
&\ & \hspace{1cm}+{1\over Y_1} \partial_y^2\theta(x-y)t_{AB}\delta^A\Psi(x)\otimes {\rm Tr}[(\delta^Bgg^{-1})T^Y]   \nonumber\\
&\ & \hspace{2cm}+{2\over Y_1} \partial_y\theta(x-y)t_{AB}\delta^A\Psi(x)\otimes \partial_y{\rm Tr}[(\delta^Bgg^{-1})T^Y]\ \Bigg).\ \ \ \   \label{gPsi}
\end{eqnarray}
The first term of the {\it r.h.s.} is reduced  to  
\begin{eqnarray}
4\pi\partial_y\theta(x-y)\Bigg(t_{AB}T^A\Psi(x)\otimes {\rm Tr}[T^B\partial_ygg^{-1}]
-{1\over Y_1}t_{AB}T^A\Psi(x) \otimes \partial_y{\rm Tr}[(\delta^Bgg^{-1})T^Y] \Bigg),  \nonumber 
\end{eqnarray}
by (\ref{linitransf}), (\ref{infini}) and (\ref{formula}).  
Plugging this expression into (\ref{gPsi}) we  find 
\begin{eqnarray}
&\ &{1\over 2\pi}\int dx\ \eta(y)\{\Psi(x)\mathop{,}^\otimes T_{++}(y)\}
 \nonumber \\
  &\ &\quad\quad  =2\Bigg(-{1\over 2}\eta(x)\partial_x \Psi(x)
  -{1\over Y_1}\partial_x\eta(x) t_{AB}T^A\Psi{\rm Tr}[(\delta^Bgg^{-1})T^Y]    \Bigg).\ \ \ \   \label{TPsi}
\end{eqnarray}
This becomes (\ref{Psiconf}) owing to a remarkable relation such that 
\begin{eqnarray}
t_{AB}T^A\Psi(x){\rm Tr}[(\delta^Bgg^{-1})T^Y] 
= \Bigg((y_a)_{N_a\times N_a}\sum_{d=1}^a N_d y_d^2\Bigg) \Psi,    \label{Remarkable}
\end{eqnarray}
with the $Y$-charge defined by (\ref{y}). It is shown as follows. We manipulate the {\it l.h.s.} by 
 (\ref{infini}), (\ref{u''}) and (\ref{u'})   as   
\begin{eqnarray}
&\ &-t_{AB}T^A\Psi{\rm Tr}[(\delta^Bgg^{-1})T^Y]  
 +t_{AB}T^A\Psi{\rm Tr}[T^BT^Y]\nonumber \\
&\ &\hspace{0.5cm}  
 = \sum_{y(\beta)>0,\ y(\gamma)>0}T_L^{-\beta}\Psi \cdot u^{\beta,-\gamma}{\rm Tr}[T_R^\gamma g^{-1}T^Yg]   
= \sum_{y(\beta)>0,\ y(\gamma)>0}T_L^{-\beta}\Psi{\rm Tr}[T_L^{-\gamma}g^{-1}T_R^\beta g]{\rm Tr}[T_R^\gamma g^{-1}T^Yg] \nonumber \\ 
&\ &\hspace{0.5cm}=  \sum_{y(\gamma)>0}t_{AB}T^A\Psi{\rm Tr}[T_L^{-\gamma}g^{-1}T^B g]{\rm Tr}[T_R^\gamma g^{-1}T^Yg]=\sum_{y(\gamma)>0}gT_L^{-\gamma}g^{-1}\Psi{\rm Tr}[T_R^\gamma g^{-1}T^Yg]\nonumber \\ 
&\ &\hspace{0.5cm}=g(t_{AB}T^A)g^{-1}\Psi{\rm Tr}[T^B g^{-1}T^Yg] - gT^Yg^{-1}\Psi{\rm Tr}[T^Y g^{-1}T^Yg].     \nonumber 
\end{eqnarray}
Here the calculation of the last two lines  was proceeded  with $g=g_Lg_H$.  
The  trace formula (\ref{trace}) reduces this relation to  
\begin{eqnarray}
t_{AB}T^A\Psi{\rm Tr}[(\delta^Bgg^{-1})T^Y]= gT^Yg^{-1}\Psi{\rm Tr}[T^Y g^{-1}T^Yg].   \label{relation2}
\end{eqnarray}
For the quantities of the {\it r.h.s.} we make the following calculations 
\begin{eqnarray}
gT^Yg^{-1}\Psi&=&
\left(
\begin{array}{c}
\noalign{\vskip0.2cm}
(gT^Y)_{\scriptscriptstyle{N_1\times N_a}}   \\
\noalign{\vskip0.2cm} 
(gT^Y)_{\scriptscriptstyle{N_2\times N_a}}   \\
\noalign{\vskip0.2cm}
 \vdots    \\
\noalign{\vskip0.2cm}
 (gT^Y)_{\scriptscriptstyle{N_a\times N_a}}     \\
\noalign{\vskip0.2cm}
\end{array}
\right)=(y_a)_{N_a\times N_a}\Psi , \nonumber\\
{\rm Tr}[gT^Y g^{-1}T^Y]&=&{\rm Tr}[g_LT^Y g_L^{-1}T^Y]={\rm Tr}[T^YT^Y]=\sum_{d=1}^a N_d y_d^2.
  \nonumber 
\end{eqnarray}
The first equation was calculated by using the form (\ref{Psi}) for $\Psi$ and (\ref{y}), while the second equation by  $g=g_Lg_H$ and (\ref{central}).
 Putting them into the {\it r.h.s.} of (\ref{relation2}) we find (\ref{Remarkable}). 

\vspace{0.5cm}
\noindent
{\bf ii. Conformal transformation of $J^\alpha_+$}

We start by calculating  the Poisson bracket using (\ref{variation})
\begin{eqnarray}
\{J_{+}^{\alpha}(x)\mathop{,}^\otimes T_{++}(y)\} =  
{\rm Tr}[\partial_x(\{g(x)\mathop{,}^\otimes T_{++}(y)\}g^{-1})gT_{R}^{\alpha}g^{-1}].    \label{JT}
\end{eqnarray}
Plug (\ref{gT'}) into the {\it r.h.s.}. Keep only the terms with $\theta(x-y)$ differentiated by $x$ because other terms drop out due to the invariance $\delta^AJ_{+}^{\beta}(x)=0$. Then (\ref{JT}) becomes  
\begin{eqnarray}
&\ &\{J_{+}^{\alpha}(x)\mathop{,}^\otimes T_{++}(y)\} \nonumber  \\
&\ &\hspace{0.5cm} = 4\pi\Bigg[\partial_x\partial_y\theta(x-y)t_{AB}{\rm Tr}[(\delta^Agg^{-1})
gT_{R}^{\alpha}g^{-1}]\otimes {\rm Tr}[(\delta^Bgg^{-1}) (\partial_ygg^{-1})] \nonumber\\
&\ & \hspace{2cm}+{1\over Y_1}\partial_x\partial_y^2\theta(x-y)t_{AB}
{\rm Tr}[(\delta^Agg^{-1})
gT_{R}^{\alpha}g^{-1}]
\otimes {\rm Tr}[(\delta^Bgg^{-1})T^Y]   \label{JT'}\\
&\ & \hspace{2cm}+{2\over Y_1} \partial_x\partial_y\theta(x-y)t_{AB}
{\rm Tr}[(\delta^Agg^{-1})
gT_{R}^{\alpha}g^{-1}]
\otimes \partial_y{\rm Tr}[(\delta^Bgg^{-1})T^Y]\ \Bigg].\ \ \ \   \nonumber
\end{eqnarray}
To simplify the expression of the {\it r.h.s.} we rewrite  the first term as
\begin{eqnarray}
&\ & 4\pi\Bigg[\partial_y\Big(\partial_x\theta(x-y)t_{AB}{\rm Tr}[(\delta^Agg^{-1})
gT_{R}^{\alpha}g^{-1}]\otimes {\rm Tr}[(\delta^Bgg^{-1}) (\partial_ygg^{-1})]\Big) \nonumber\\
&\ &\hspace{1cm} -\partial_x\theta(x-y)t_{AB}{\rm Tr}[(\delta^Agg^{-1})
gT_{R}^{\alpha}g^{-1}]\otimes \partial_y{\rm Tr}[T^B (\partial_ygg^{-1})]   \nonumber\\
&\ &\hspace{1cm} +\partial_x\theta(x-y)t_{AB}{\rm Tr}[(\delta^Agg^{-1})
gT_{R}^{\alpha}g^{-1}]\otimes{1\over Y_1}\partial_y^2{\rm Tr}[(\delta^Bgg^{-1})T^Y] \Bigg],  \label{a}
\end{eqnarray}
by (\ref{infini}) and (\ref{formula}). 
Plug this result into (\ref{JT'}).  Multiplying $\eta(y)$ on both sides we integrate (\ref{JT'}) over $y$. The integration is  reduced to 
\begin{eqnarray}
&\ & {1\over 2\pi}\int dy\eta(y)\{J_{+}^{\alpha}(x)\mathop{,}^\otimes T_{++}(y)\} 
      \nonumber\\
&\ &\hspace{1cm} =-\eta(x)\partial_xJ_+^\alpha 
- 2\partial_x\eta\Big(t_{AB}{\rm Tr}[(\delta^Agg^{-1})
gT_{R}^{\alpha}g^{-1}] {\rm Tr}[(\delta^Bgg^{-1}) (\partial_xgg^{-1})]\Big) \nonumber\\
 &\ &\hspace{4.4cm}\ \ +2\partial_x^2\eta\Big({1\over Y_1}t_{AB}
{\rm Tr}[(\delta^Agg^{-1})
gT_{R}^{\alpha}g^{-1}]
 {\rm Tr}[(\delta^Bgg^{-1})T^Y]\Big).  
 \label{JT''}
\end{eqnarray}
The first term of the {\it r.h.s.} was obtained from the second term in (\ref{a}) by using the trace formula (\ref{trace}) and (\ref{variation}). 
The second and third terms  are respectively vanishing as follows. For the third term we have 
\begin{eqnarray}
&\ &t_{AB}{\rm Tr}[(\delta^Agg^{-1})gT_{R}^{\alpha}g^{-1}]{\rm Tr}[(\delta^Bgg^{-1})T^Y]  \nonumber \\
&\ &\hspace{1cm} = {1\over 2}{\rm Tr}[gT_{R}^{\alpha}g^{-1}T^Y]  -\hspace{-0.3cm} \sum_{y(\beta)>0,\ y(\gamma)>0}{\rm Tr}[T_L^{-\beta}gT_{R}^{\alpha}g^{-1}]u^{\beta,-\gamma}     {\rm Tr}[T_R^\gamma g^{-1}T^Yg],   \label{3rd} 
\end{eqnarray}
by (\ref{infini}) and (\ref{u'}).  The sum over $\beta$ of the second term in the {\it r.h.s.} is calculated as 
\begin{eqnarray}
\sum_{y(\beta)>0}{\rm Tr}[T_L^{-\beta}gT_{R}^{\alpha}g^{-1}]u^{\beta,-\gamma}&=&\sum_{y(\beta)>0}{\rm Tr}[T_L^{-\beta}gT_{R}^{\alpha}g^{-1}]
{\rm Tr}[T_L^{-\gamma}g^{-1}T_R^\beta g]  \nonumber\\
&=&t_{AB}{\rm Tr}[T^AgT_{R}^{\alpha}g^{-1}]
{\rm Tr}[T_L^{-\gamma}g^{-1}T^B g]={1\over 2}\delta^{\alpha,-\gamma},
\end{eqnarray}
by means of (\ref{u''}) and the trace formula (\ref{trace}).  
Due to this formula the {\it r.h.s.} of (\ref{3rd}) vanishes. So does the third term of the {\it r.h.s.} in (\ref{JT''}). Similarly we can show that the second term  in (\ref{JT''}) is vanishing.  Thus the conformal transformation (\ref{J+}) was shown.

\section{Conclusions}

In this paper we have given a general account of constrained WZWN models on $G/\{S\otimes U(1)^n\}$ to the most extent. We have found the $G$-primary in an arbitrary representation of $G$, which satisfies the classical exchange algebra. It was shown to have conformal weight given by (\ref{Psiconf}) with respect to $T_{++}$. The constraints were found to take the specific form as (\ref{constraint}). These were imposed so that the constrained currents  are invariant under the gauge-fixed transformation (\ref{transf}), i.e., 
$\delta J^\alpha_+=0$. This was important for the self-consistency of the constraints. The constrained currents should transform as  conformal primaries of weight $0$.  This is also important for the consistency of the constraints. Otherwise the constraints would break conformal symmetry.  It was checked a posteriori.

To conclude this paper we would like to  make comments on the classical r-matrix
 which we encounter in studying  two-dimensional non-linear $\sigma$-models on the symmetric space $G/H$ and the principal chiral coset space $G\otimes G/G$. There exists a  current $j(x)$ valued in the Lie algebra of $G$ which 
 satisfies $d*j=0$ and $dj+j\wedge j=0$. From such a current  we may construct a one-parameter family of a flat connection $L(\lambda,x)$, called  Lax operator\cite{Bena}. The Poisson brackets for $j(x)$ are set up on the $t=const$ plane. 
By using them 
$L(\lambda,x)$ was shown to satisfy the Poisson algebra 
\begin{eqnarray}
\{L(\lambda,x)\mathop{,}^\otimes L(\mu,y)\}=[r_{\lambda\mu},L(\lambda,x)\otimes 1+1\otimes L(\mu,y)]\delta(x-y) \label{LL}
\end{eqnarray}
or some modified one\cite{Maillet}. We may also be interested in  a path-ordered exponential of the Lax operator 
$$
T(\lambda)=P\exp \int_{-\infty}^\infty dx\ L(\lambda, x),
$$
called monodromy matrix. By similar calculations  it was shown that $T(\lambda)$ satisfies the quadratic algebra\cite{Maillet, Bernard, Magro}  
\begin{eqnarray}
\{T(\lambda)\mathop{,}^\otimes T(\mu)\}= [r_{\lambda\mu},T(\lambda)\otimes T(\mu)].   \label{TTcom}
\end{eqnarray}
The algebra (\ref{TTcom}) is a classical exchange algebra of the same type as (\ref{CEA'}), as can be seen by putting (\ref{CEA'}) in the form (\ref{CEA}) with the r-matrix  (\ref{rxy}). The appearance of a commutator on the {\it r.h.s.}  is due to the tensor property of $T(\lambda)$, i.e., if  the r-matrix is given in the representation  $\mbox{\boldmath $N$}$, it acts on $T(\lambda)$ as $\mbox{\boldmath $N$}\otimes \overline{\mbox{\boldmath $N$}}$, but on $\Psi(x)$ as $\mbox{\boldmath $N$}$. The r-matrix $r_{\lambda\mu}$ in (\ref{LL}) and (\ref{TTcom})  was discussed in \cite{Maillet,Bernard,Magro}  to find an explicit form as a function of parameters $\lambda$ and $\mu$. On the contrary  the r-matrix $r_{xy}$, given by (\ref{rxy}) with (\ref{r-matrix}), does not have such parameter dependence, but a position dependence  on the light-like plane through $\theta(x-y)$. So our exchange algebra (\ref{CEA'}) is still different from both of (\ref{LL}) and (\ref{TTcom}). 
 These arguments about non-linear $\sigma$-models on $G/H$ were also extended by taking $G$ to be the supergroup $PSU(2,2|4)$ or its variants\cite{Magro, Berk}. Then there appears a two-dimensional topological term which descends  from the WZ term. In that case the Poisson structure is modified by  first-class constraints such as one encountered in the covariant formalism of the Green-Schwarz superstring. A careful study is needed for disentangling them from  second-class constraints.





\bibliographystyle{model1-num-names}
\bibliography{<your-bib-database>}



\end{document}